\documentclass[default,iicol]{sn-jnl}

\usepackage{graphicx}%
\usepackage{multirow}%
\usepackage{amsmath,amssymb,amsfonts}%
\usepackage{amsthm}%
\usepackage{mathrsfs}%
\usepackage[title]{appendix}%
\usepackage{xcolor}%
\usepackage{textcomp}%
\usepackage{manyfoot}%
\usepackage{booktabs}%
\usepackage{algorithm}%
\usepackage{algorithmicx}%
\usepackage{algpseudocode}%
\usepackage{listings}%
\usepackage{placeins}

\usepackage[sorting=none]{biblatex} 
\begin{document}

\newcommand{\hww}[0]{$H\rightarrow WW^{*}$~}

\title[Article Title]{Isolating semi-leptonic \hww decays for Bell inequality tests}


\author[1,2,3]{\fnm{Federica} \sur{Fabbri}}\email{federica.fabbri@glasgow.ac.uk}
\author[1]{\fnm{James}    \sur{Howarth}}\email{james.howarth@glasgow.ac.uk}
\author[1]{\fnm{Th\'eo}     \sur{Maurin}}\email{t.maurin.1@research.gla.ac.uk}

\equalcont{These authors contributed equally to this work and are listed in alphabetical order.}

\affil[1]{\orgdiv{SUPA - School of Physics and Astronomy}, \orgname{University of Glasgow}}

\affil[2]{\orgdiv{Dipartimento di Fisica e Astronomia A. Righi}, \orgname{Universitá di Bologna}}
\affil[3]{\orgname{INFN Sezione di Bologna}}


\abstract{We present a method for identifying $H\rightarrow WW^* \rightarrow \ell \nu j j$ events in the presence of large Standard Model backgrounds and illustrate how this decay mode may be applied to the study of Bell-type Inequalities. Our findings reveal the feasibility of complete reconstruction of such Higgs decays and the efficacy of our suggested combination of selection criteria in effectively mitigating the otherwise overwhelming backgrounds. Our approach is based on a combination of bottom and charm tagging, alongside innovative reconstruction techniques. 
A realistic simulation based on publicly available object identification, reconstruction, and tagging efficiencies from the ATLAS experiment is used to explore the potential sensitivity to violations of the Collins-Gisin-Linden-Massar-Popescu (CGLMP) inequality in existing and expected future data collected at the Large Hadron Collider (LHC). It is found that, the proposed method provides a powerful means of distinguishing the Higgs decay mode from the background, allowing us to achieve an expectation of approximately 3$\sigma$ significance in detecting violations of these inequalities with 300 $\mbox{fb}^{-1}$ of data, soon-to-be collected by the LHC.}


\maketitle

\section{Introduction}\label{sec:info}

The production of Higgs bosons decaying to a pair of $W$ bosons has been observed by the ATLAS  and CMS collaborations at $\sqrt{s} = 8$ TeV, and more recently has been probed at $\sqrt{s} = 13$ TeV by both collaborations. The total cross-section is measured to a precision of 12\%/12\% in the gluon-gluon fusion production mode and to 25\%/39\% in the associated vector boson production modes by ATLAS and CMS, respectively \cite{ATLAS:2022ooq,CMS:2022uhn}. However, in each of these results, only the fully leptonic decay modes ($H\rightarrow WW^*\rightarrow \ell \nu \ell \nu$, where $\ell = e,\mu$\footnote{For the duration of this article, charge conjugation is assumed unless explicitly stated otherwise. That is, $W^{+}$ and $W^{-}$ will both be referred to as $W$, similarly for leptons and quarks.}) were considered due the significant Standard Model (SM) background processes that mimic this final state when considering cases where one or both $W$'s may decay hadronically. Due to the relatively low leptonic branching fraction of the $W$, these results have large statistical uncertainties, at the same level of the experimental and theoretical systematic uncertainties. Combining the currents results with events collected from final states not restricted to be fully leptonic would more than double the statistics available for the $H\rightarrow WW^{*}$ measurements. Another severe limitation of the dileptonic final states is the presence of two neutrinos, that makes high resolution reconstruction of the Higgs boson kinematics almost impossible. This limits the potential to measure phenomena unique to the Higgs boson with LHC data, such as a proposed measurement of quantum entanglement in a qutrit system \cite{Barr:2021zcp}. In this article, we propose a method to identify and fully reconstruct the more prevalent semi-leptonic decay mode ($H\rightarrow WW^* \rightarrow \ell \nu j j$, where $j=q=c,s$) whilst also reducing the SM background to a manageable level. Furthermore, we demonstrate the utility of these techniques by investigating the potential to measure Bell Inequality violation in Higgs boson decays using a combination of data already collected at the LHC experiments during Run2 ($\sqrt{s} = 13$ TeV) and data that are being collected in the emergent Run3 ($\sqrt{s} = 13.6$ TeV) and the future High Luminosity LHC (HL-LHC).

\section{Background Suppression and Event Reconstruction}
\label{sec:background_suppression}

In the \hww decay mode, one of the $W$ bosons is necessarily off its mass-shell (indicated by an asterisk) due to the mass of the Higgs boson ($\sim 125$~GeV) being significantly lower than twice the on-shell $W$ mass ($\sim 160$~GeV). All studies of the $H\rightarrow WW^{*}$ process have so-far focused on the fully leptonic final state ($\ell \nu \ell \nu$) since this is the only decay mode that is not overwhelmed by background events from other SM processes.
In particular, the semi-leptonic decay mode ($H\rightarrow WW^{*} \rightarrow \ell \nu q q'$) is not normally considered as it is dwarfed by the $W$+jets process. The cross-section of semi-leptonic \hww production is $4.3$~pb ($m_H = 125.0$~GeV) \cite{pdg,HYR4} but the cross-section of the $W(\rightarrow l\nu)$+jets process is $\sim 42000$~pb at next-to-leading-order accuracy in perturbative QCD\cite{madgraph}. In order to successfully isolate the Higgs signal, the $W$+jets process must be suppressed by 4 orders of magnitude.

The study of this specific final state offers advantages beyond the mere addition of a new channel to Higgs measurements, as it facilitates the full reconstruction of the Higgs system, a task that proves challenging in fully-leptonic \hww decays.

In hadron collider experiments, neutrinos are not measured directly but their presence is inferred by requiring momentum conservation in the plane transverse to the beam direction. It is possible to determine the total $p_x$ and $p_y$ components of the sum of all neutrinos present in the event via missing transverse momentum ($P_{x/y}^{miss}$), but not their $p_z$ or, in the case of multiple neutrinos, their individual $p_x$ or $p_y$ components. This presents a problem for the measurement of the properties of the Higgs boson, such as the measurement of Bell inequalities proposed by Barr~\cite{Barr:2021zcp}, because these require that the Higgs itself, and all of its decay particles, be fully reconstructed. At the time of writing, no dedicated studies exist on how to solve this problem for the fully-leptonic \hww case. In the semi-leptonic case, if the leptonically-decaying $W$ were assumed to be on-shell (and therefore, the hadronic $W$ to be off-shell), the on-shell $W$ mass could be used as an additional constraint, along with the $P_{x/y}^{miss}$, to reconstruct the leptonic $W$. Two additional hadronic jets in an event would then need to be paired under some criterion (such as $m(jj) < 80$~GeV) to reconstruct the off-shell hadronic $W$ and both would need to combine to reconstruct a Higgs boson with a mass of 125 GeV. This approach would be similar to the techniques employed in the reconstruction of semi-leptonic $t\bar{t}$ decays~\cite{ATLAS_top_ljets}. However, whilst this approach would have an high efficiency on the \hww signal, it would also select a large fraction of $W$+jets, which also contain an on-shell leptonic $W$ and could easily have additional hadronic jets that satisfy whatever criterion were chosen for the Higgs reconstruction. The problem of isolating the signal Higgs with respect to the $W$+jets background would remain.

We propose instead to sacrifice using the $P_{x/y}^{miss}$ directly and the requirement that the leptonic $W$ be on-shell and impose a stricter requirement on the other side of the event, i.e. that the hadronically decaying $W$ boson be on-shell. We then use a technique adapted from $t\bar{t}$ dileptonic final state reconstruction called Neutrino Weighting~\cite{D0:1997pjc} to reconstruct the final state and, additionally, to also suppress the SM $W$+jets background.

\subsection{Monte Carlo Processes and Detector Effects}

To determine the feasibility of differentiating signal \hww events from background SM processes and of observing Bell's inequality violation in Higgs events under real experimental conditions, signal and background events are simulated using the \textsc{Powheg} generator~\cite{Powheg1,Powheg2,Powheg3,Powheg4}. One million signal events are generated for semi-leptonically decaying \hww events via the gluon-gluon fusion production mode \cite{PowhegHiggs}. Similarly, one million background events each are generated for $t\bar{t}$ , $W$+jets and $tW$ processes.  In all cases, \textsc{Pythia}~\cite{Pythia82,Pythia83} is used to perform the parton shower and hadronisation of the events. The only exception are the Diboson processes, generated using \textsc{Sherpa}~\cite{Sherpa:2019gpd}. Only processes that generate a single lepton in the final state are considered.

RIVET~\cite{rivet} is used to process the HEPMC~\cite{hepmc2,hepmc3} events and reconstruct physics objects. Physics objects are constructed from stable particles in the event record (particles with a mean lifetime greater than 30 ps). Electrons and muons are identified as originating from a $W$ boson decay (including via intermediate $\tau$ leptons) and all photon radiation within a cone of $\Delta R < 0.1$ (where $\Delta R^2 = \Delta \eta^2 + \Delta \phi^2)$ is summed into the lepton four-momentum and removed from the event record (excluding photons from hadrons decays). Jets are reconstructed using the anti-$k_{t}$ algorithm with a distance parameter of $R=0.4$ and are tagged as having $b$ or $c$ flavour via ghost matching~\cite{antikt}. $P_{x/y}^{miss}$ is defined using only neutrinos originating from a $W$ boson decay (including via intermediate $\tau$ leptons). Neutrinos and leptons deriving from hadron decays are included in the input to the jet clustering. These object definitions follow those typically used by the ATLAS and CMS collaborations during LHC Run2. 

In order to simulate the effect of the detector response, publicly available acceptance and efficiency values from the ATLAS collaboration are applied to the physics objects.
The trigger efficiencies for electrons~\cite{ATLASTriggerEgamma} and muons~\cite{ATLASTriggerMuon} are based on the assumption of using single lepton unprescaled triggers (with a $p_T$ threshold of 20 GeV). The major effect of these efficiencies is to effectively set the minimum $p_T$ requirement for the leptons employed in the analysis. The lepton identification and reconstruction efficiencies for electrons and muons are taken to be 90\% and 96.1\%, respectively, based on the \emph{medium} definitions used by ATLAS~\cite{ATLASElectronID,ATLASMuonReco}. We assume a jet reconstruction efficiency of unity and neglect the effects due to limited resolution. Jet smearing based on \cite{ATLAS_jet_smearing} was tested and was found to have a limited impact on the results described in the following sections. Furthermore, recent experimental results that demonstrate the significant potential for improving jet resolution and response through the utilization of machine learning techniques~\cite{ATLASJES} imply that these effects will be subdominant in analyses on real data. Moreover, in real-experiment analyses, the impact of jet resolution and response can be further mitigated by implementing dedicated regions and observables designed to constrain the jet calibration in-situ. 

\subsection{Neutrino Weighting}

\begin{figure}
\includegraphics[width=\columnwidth]{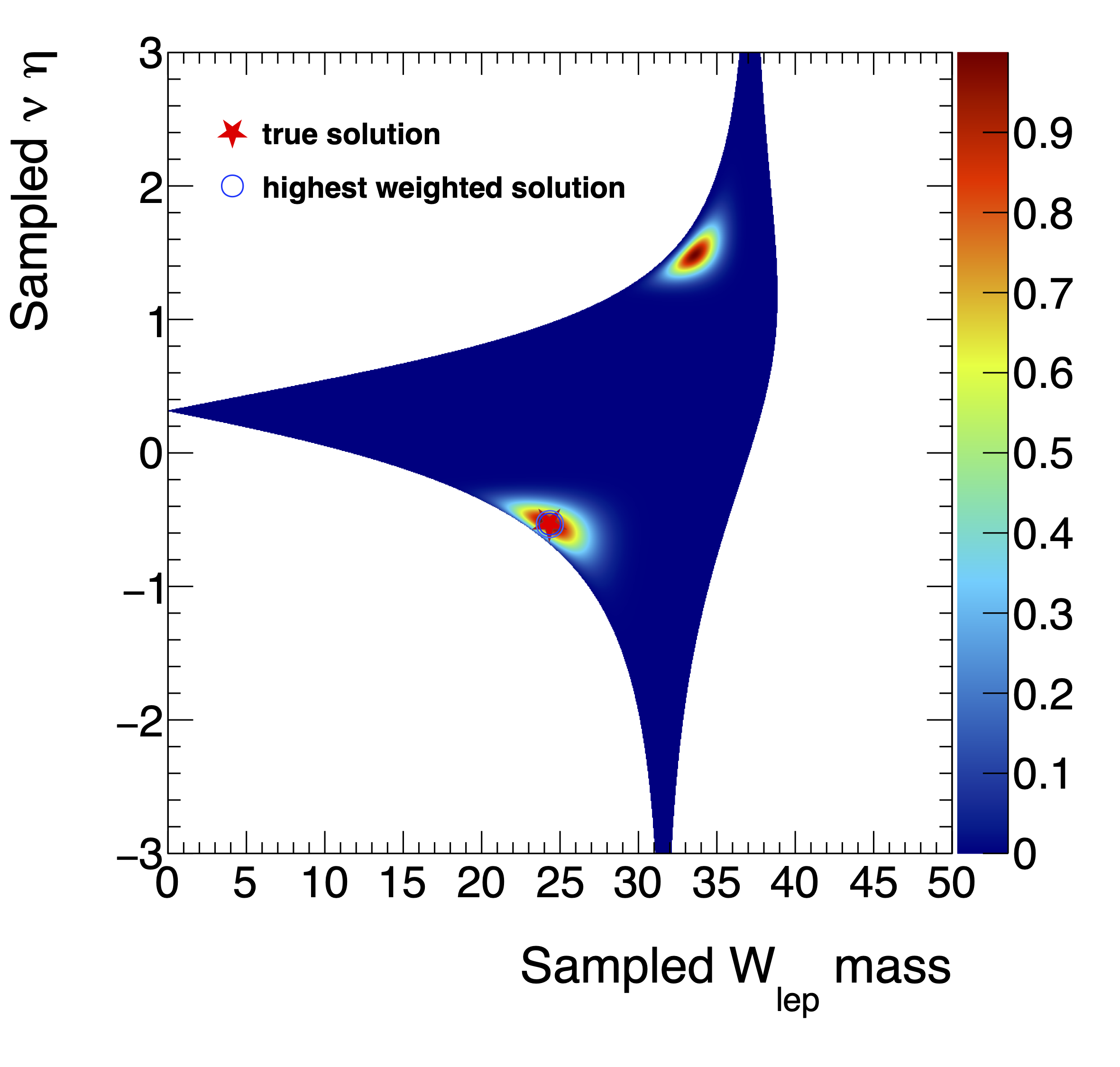}
\caption{The distribution of weights for each assumption of leptonic $W$ boson mass and neutrino pseudo-rapidity, with markers indicating the location of the true values of these parameters and the scan point with the highest weight.}
\label{fig:NW}
\end{figure}

Neutrino Weighting (NW) is a technique used to reconstruct leptonically decaying $W$ bosons. It has been used to great effect in the dileptonic $t\bar{t}$ signature, most recently in measurements of spin correlation in $t\bar{t}$ events by ATLAS~\cite{ATLAS_spin,NW}, but it requires some adaptation to be applicable to the $H\rightarrow WW^{*} \rightarrow \ell \nu j j$ signature. NW solves the problem of reconstructing $W$ bosons that decay leptonically by introducing assumptions on the unconstrained parameters. In the case of dileptonic $t\bar{t}$, these assumptions are the values for the pseudo-rapidity of each of the two neutrinos in the final state. In our modification, these assumptions are the pseudo-rapidity of the neutrino ($\eta_{\nu}$) and the off-shell mass of the leptonically decaying $W$ boson ($m_{W(lep)}$). Since neither of these values are known a-priori, the full system must be reconstructed under many possible assumptions for $m_{W(lep)}$ and $\eta_{\nu}$. With these two assumptions, as well as an assumption on the mass of the Higgs (which we take to always be 125 GeV) and the reconstructed mass of the hadronically decaying $W$, the system is fully constrained and can be solved for the  neutrino's momentum $x$ and $y$ components\footnote{The full details of this calculation and a python implementation are available from the authors upon request.}. There are zero, one, or two potential real solutions (corresponding to the roots of the polynomial) for each assumption of $m_{W(lep)}$ and $\eta_{\nu}$. For each assumption, a weight is generated based on the difference between the observed $P_{x/y}^{miss}$ and the reconstructed neutrino kinematics:

\begin{equation}
w = \exp(\frac{(\nu_x - P_x^{miss})^2}{\sigma_x^2}) \cdot \exp(\frac{(\nu_y - P_y^{miss})^2}{\sigma_y^2})\mbox{,}    
\label{eq:weight}
\end{equation}

\noindent where $\nu_{x/y}$ is the $x/y$ component of the reconstructed neutrino's momentum, $P_{x/y}^{miss}$ is the observed $P_{x/y}^{miss}$ in the experiment in the $x/y$ direction, and $\sigma_{x/y}$ is the experimental resolution of $P_{x/y}^{miss}$ (the values for these resolutions only serve to scale the weight and have no physical impact on the result if taken equal in $x$ and $y$). The resultant weight is high when the reconstructed neutrino matches the observed $P_{x/y}^{miss}$ and low when it does not and reaches a maximum when the chosen values for $\eta_\nu$ and $m_{W}$ are closest to the true ones. An example of the distribution of weights for a single example collision event is shown in Fig.~\ref{fig:NW}. As can be seen from Eq.~\ref{eq:weight}, the weight is defined between 0 and 1, and the distribution peaks at the same values as the true $m_{W(lep)}$ and $\eta_{\nu}$. One can also observe a second peak, unrelated to the true values of $m_{W(lep)}$ and $\eta_{\nu}$ corresponding to the second root. In the example event shown in Fig.~\ref{fig:NW}, the solution with the highest weight lies at the same values as the true quantities. A discussion of the performance of the NW algorithm is reported in Section~\ref{sec:event_selection}.

The weight from NW functions is more than just a method of selecting the most-likely values of $m_{W(lep)}$ and $\eta_{\nu}$, it can also be used to suppress background processes with remarkable efficiency. This is discussed further in Section~\ref{sec:event_selection}.

\subsection{Charm Tagging}
\label{sec:charm_tagging}

Thanks to the enormous statistics provided by the LHC, it is possible to restrict the decay signature to events where the hadronic $W$ boson decays to a charm and strange\footnote{To the aim of the analysis reported here the exact flavour of the quark produced in association with the $c$ is not relevant, given that all down-type quark have the same spin analyzing power.} quark. As we have already assumed one on-shell $W$ bosons decaying hadronically to perform the NW reconstruction, the major backgrounds to the \hww signal arise from the production of a single leptonically decaying $W$ boson with additional jets that arise from gluon radiation, from pairs of $W$ bosons where only one boson decays hadronically and from $t\bar{t}$ events, that contain two $W$ bosons originating by the top quark decay. 

Both the ATLAS and CMS experiments heavily utilise sophisticated algorithms to identify (tag) hadronic jets containing $B$-hadrons, which experiments use to infer the presence of $b$ quarks in an event. The predominant feature used by these algorithms is the long lifetime of $B$-hadrons, leading to so called \emph{displaced vertices} in the experiment inner trackers, a feature conspicuously absent from jets originating from so-called \emph{light} jets (i.e. those initiated by gluons or $u$, $d$, $s$ quarks). Jets originating from charm quarks occupy an interesting grey area as hadrons with charm flavour also have sufficiently long lifetimes to create displaced vertices and both ATLAS and CMS have created dedicated $c$ tagging algorithms to identify them. This can be useful for reducing the $W$+jets background to our signal. The additional jets are more likely to be light jets from gluons or the light flavour quarks (such as $u\bar{u}$ or $d\bar{d}$) compared to heavier quarks pairs (such as $c\bar{c}$), whereas in our signal, the on-shell hadronic $W$ decays to a $ud$ pair or a $cs$ pair with similar branching fractions. By requiring the presence of exactly one $c$-tagged jet, the SM background (which contains either none or two) can be reduced.

In this study, we simulate charm tagging using a technique called `ghost-matching', where jets in Monte Carlo events are determined to have either $b$ or $c$ flavour by setting unstable B and D hadrons energies to zero and allowing them to be clustered into stable particle jets. Any jet with a clustered B hadron is determined to be a $b$-jet and any with a clustered D hadron and no B hadrons is tagged as a charm jet. Only hadrons with $p_{T}$ above 5~GeV are considered. We then apply efficiency and mis-tag factors based on recent ATLAS values that are summarised in Table~\ref{tab:ctagging_ATLAS}\cite{ATLAS_btag,ATLAS:2022qxm}.

\begin{table}
\caption{$c$ and $b$-tagging efficiencies used in the ATLAS experiment~\cite{ATLAS_btag,ATLAS:2022qxm}. The numbers refer to the probability for the tagging algorithms to assign the $c$ or $b$ flavour to jets containing a $B$-hadron ($b$-jet), a $c$-hadron and no $B$-hadrons ($c$-jet) or no heavy-flavour hadrons ($l$-jet). These numbers are directly used in the analysis presented in this paper to emulate realistic $c$-tagging performances.}
\centering
\begin{tabular}{c c c}
\toprule
true flavour  &  $c$-tagging efficiency & $b$-tagging efficiency   \\
\midrule
$b$-jet           &  0.14  &  0.77 \\ 
$c$-jet           &  0.4   &  0.2 \\ 
$l$-jet           &  0.016 &  0.008 \\  
\bottomrule
\end{tabular}
\label{tab:ctagging_ATLAS}
\end{table}

For the $H\rightarrow WW^{*} \rightarrow c j\ell \nu $ we are particularly interested in identifying the following features:
\begin{itemize}
    \item Exactly one $c$-tagged jet.
    \item One or more light jets (additional light jets could be radiated in the event).
    \item Exactly zero $b$-tagged jets.
\end{itemize}
The first two points are derived from the condition that one of the $W$ bosons decays hadronically to a $cs$ pair. The third point is necessary to reject the $t\bar{t}$ background process, which contains two $b$ quarks in the final state. The second point requires the least consideration; all algorithms employed by LHC experiments are very efficient at rejecting light jets, on the other hand the interplay between $b$-tagging and $c$-tagging is non-trivial. The efficiency to mis-tag a $b$-jet for a $c$-jet, and vice-versa, can be relevant, as shown in Table~\ref{tab:ctagging_ATLAS}. 

\subsection{Event Selection}
\label{sec:event_selection}

\begin{table*}
\centering
\caption{The expected yields for an integrated luminosity of 300 fb$^{-1}$ in a selection with idealised charm tagging and Run2-like charm tagging. The signal over the signal plus background (S/(S+B)) is also shown. The uncertainties are statistical only.}
\label{tab:yields}
\begin{tabular}{c r c l r c l}
\toprule
Process  &\multicolumn{3}{c}{idealised} & \multicolumn{3}{c}{$\epsilon_c=40\%$} \\ 
\midrule

$W$ + jets          & 13131 &$\pm$& 785 &  10444 &$\pm$& 664 \\
$WW$                &  2298 &$\pm$&  31 &   1137 &$\pm$&  22 \\
$t\bar{t}$          &   601 &$\pm$&  76 &   1453 &$\pm$& 119 \\
$tW$                &   217 &$\pm$&   8 &    350 &$\pm$&  11 \\
Higgs               &  5967 &$\pm$&  76 &   2843 &$\pm$&  56 \\
\midrule
S/(S+B)             & \multicolumn{3}{c}{0.27} & \multicolumn{3}{c}{0.18} \\
\bottomrule
\end{tabular}

\end{table*}

Based on the discussions in Sections~\ref{sec:charm_tagging}, the following criteria are used to select events:
\begin{itemize}
    \item pre-selection:
    {\begin{itemize}
    \item Exactly 1 lepton with $p_{T} >$ 20 GeV
    \item Exactly 0 $b$-tagged jets
    \end{itemize}}
    \item $c$-tagging selection:
    {\begin{itemize}
    \item 2 or more jets, exactly one of which must be $c$-tagged.
    \item At least 1 ($c$-jet,$l$-jet) pair with $|m_{cl} - 80.6| < 10$~GeV
    \end{itemize}}
    \item A reconstructed leptonic $W$ boson from NW with $w > 0.7$.
    \item Maximum 2 light jets.
    \item Invariant mass of the lepton and the $c$-tagged jet $m(\ell c) <$ 80~GeV.
\end{itemize}

If there are multiple combinations of light-jets and $c$-jet, the pair with mass closest to the $m_W=80.6$ GeV  is chosen as reconstructed hadronic $W$ and employed in the NW.
Using this selection we consider two charm tagging possibilities; an idealised case (with perfect identification efficiency and perfect $b$ and light jet rejection) and a realistic case with $\epsilon_c=40\%$, shown in Table~\ref{tab:ctagging_ATLAS}. The latter is representative of the possibility with state-of-the-art charm tagging, whereas the former is provided as an indication of the best-case-scenario if significant improvements could be made. In all of the following tables and figures the realistic charm tagging case is used unless otherwise specified. 

Given that the mass of the hadronic $W$ is an ingredient of the NW, the charm tagging also slightly influences the performance of the NW. The algorithm has indeed a 78$\%$ probability of finding a solution in signal events in the idealised case, while the probability is 72$\%$ in the realistic case. The resolution of $m_{W}$ mass reconstruction is slightly below 10$\%$ in the idealised case and slightly above in the realistic one. The performance of the NW can be further optimised by fine tuning the number of points scanned in the $\eta_\nu$ and $m_{W}$ plane.

\begin{figure}
     \centering
     \includegraphics[width=0.48\textwidth]{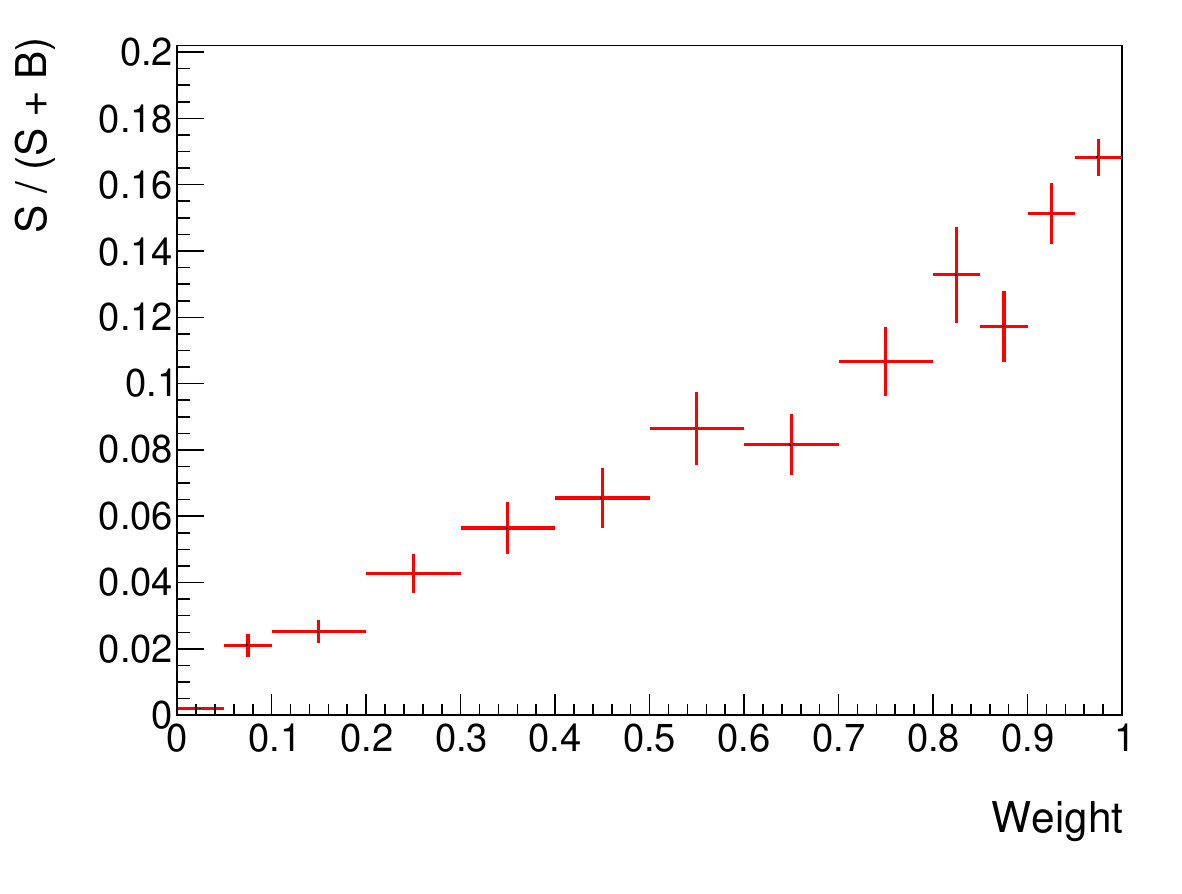}
    \hfill
    \caption{The signal over signal plus background as a function of the requirement applied on the score from the Neutrino Weighting. Only events passing the pre-selection and the $c$-tagging selection are included in the plot.}
     \label{fig:soverb}
\end{figure}

\begin{figure}
     \centering
     \includegraphics[width=0.48\textwidth]{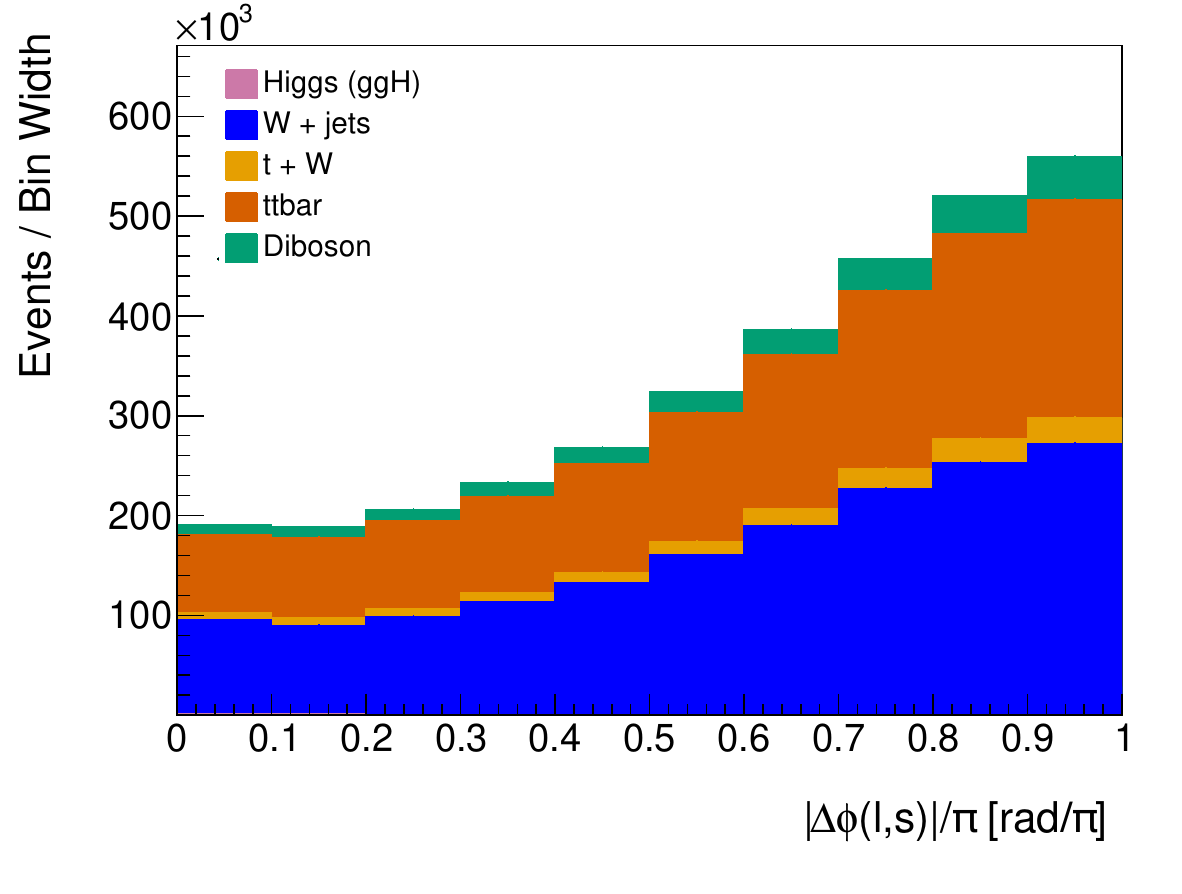}
     \hfill
     \includegraphics[width=0.48\textwidth]{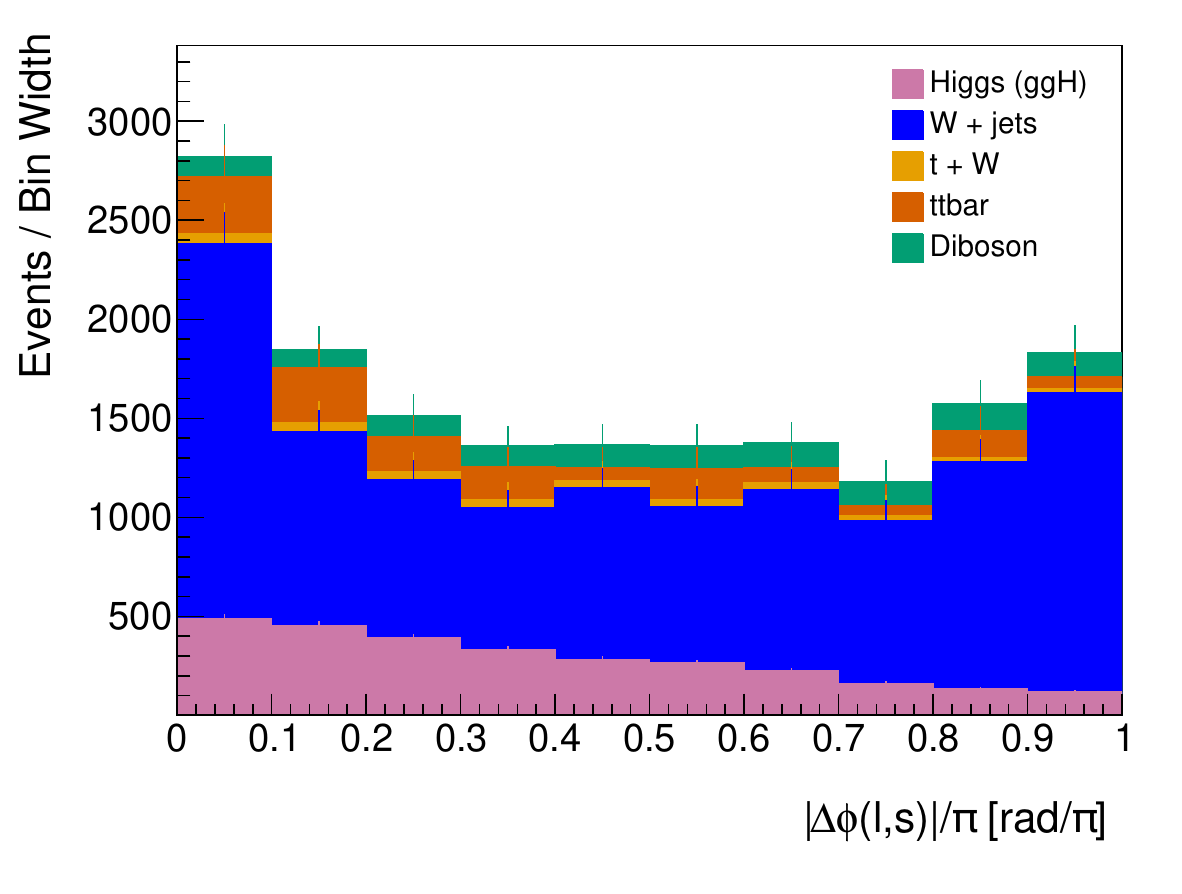}
    \hfill
     \caption{Number of events as a function of the azimuthal angle between the lepton and the down-type quark ($\Delta\phi(l,s)$) applying only the pre-selection and $c$-tagging requirements (top) and applying the whole selection (bottom). }
     \label{fig:selection_mls}
\end{figure}

The expected fraction of the signal (S) compared to the full simulation, including backgrounds (B), as a function of the weight from NW is presented in Fig.~\ref{fig:soverb}. Here the NW requirement is applied on top of the pre-selection and $c$-tagging requirements. Applying only the pre-selection, the fraction of signal over the entire simulated sample is $\sim$0.0001. The $c$-tagging requirements help in selecting the signal, increasing the $\frac{S}{S+B}$ by a factor ten, but it is only the constraint on the NW score that allows to really isolate the signal. The requirement at 0.7 on the NW score has indeed a 0.45 efficiency on the signal, against a 0.005 efficiency on the backgrounds.
On top of the NW score request two additional selection requirements are applied, mainly targeting the reduction of $t\bar{t}$ background. To enter the selection the two $b$-jets included in the $t\bar{t}$ process must be mis-reconstructed as light jets or $c$-tagged jets. The limit on the number of jets remove events in the first category while selecting on $m(\ell c)$ removes events from the second category. The $m(\ell c)$ distribution is indeed related to the mass of the parent particle, so it shows a very distinct distribution in the $t\bar{t}$ case, where the lepton and the $c$-tagged jet (mis-reconstructed $b$-jet) derive from a top-quark decay.
The expected composition of the data after the pre-selection and $c$-tagging requirements and after the whole selection is presented in Fig.~\ref{fig:selection_mls}. Even with the requirements on the hadronic $W$ mass and the veto on $b$-tagged jets, the $W$+jets and $t\bar{t}$ processes significantly dominate over the signal before the requirement on the weight from NW is applied. After the whole selection, the contribution from $t\bar{t}$ and the $W$+jets processes is drastically reduced and the signal is well visible.

The expected number of signal and background events for 300 fb$^{-1}$ of $\sqrt{s}=$ 13 TeV data (the total expected Run2 + Run3 luminosities for ATLAS and CMS) are presented in Table~\ref{tab:yields}. 2843 signal events are expected under realistic charm-tagging conditions, with a signal purity of 18\%. This is a remarkably high purity given that the cross-section for the background process ($W$+jets in particular) are orders of magnitude higher than the signal process. Under perfect charm-tagging performance, these numbers increase to 5967 signal events and a purity of 27\%. Though such performance is unlikely to be achievable, these values serve to act as an upper bound on the performance of the techniques described in this paper. When including realistic jet smearing the purity degrades to 13\% for the realistic charm tagging case and this is likely a fair representation of what is achieve-able under LHC Run2 conditions. 
However, advanced analysis techniques, such as multivariate approaches, could be employed to maximise the separation between signal and background exploiting other observables that manifest a different behavior between the signal and the main backgrounds, e.g. $\Delta\phi(\ell ,s)$ shown in Fig.~\ref{fig:selection_mls} (bottom). 


\section{Observing Bell-Inequality Violation in Higgs Events}

Many Bell-type inequalities exist for a variety of different scenarios. For the \hww process, which is a pair of three-outcome spin states (qutrit), the Collins-Gisin-Linden-Massar-Popescu (CGLMP) inequality has been shown to be the optimal choice~\cite{CGLMP1,CGLMP2}. The applicability of this inequality to fully-leptonic \hww events has already been discussed in previous works. We follow the strategy proposed by Barr~\cite{Barr:2021zcp} with modifications only to account for the semi-leptonic final state. This is possible because the spin analysing power of the down-type quarks deriving by the W decay is 1 as in the charged leptons case. As a consequence the technique introduced in~\cite{Barr:2021zcp} can be applied also to the case presented here using the lepton and the down-type quark instead of the charged lepton pair to reconstruct the observables of interest. The down-type quark employed in the following is identified with the light jet used to reconstruct the hadronic $W$.

The CGLMP inequality ($\mathcal{I}_{3}$) may be constructed from the expectation values of operators ($\mathcal{B}$) constructed from angular observables based upon a choice of three orthonormal axes ($\hat{x}\hat{y}\hat{z}$). These axes are defined based on the direction of the $W^{+}$ boson momenta ($\hat{x}$) in the $W^{+}W^{-}$ rest frame, the direction perpendicular to the plane formed by $\hat{x}$ and the direction of the incoming protons ($\hat{y}$), and the right-handed remaining orthogonal direction ($\hat{z}$) (a more in-depth discussion on these definitions and their similarity to those used in top spin measurements is provided in ~\cite{Barr:2021zcp}). The CHLMP inequality is therefore defined as:
\begin{equation}
    \mathcal{I}^{xyz}_{3} = \max(\langle \mathcal{B}^{xy}_{\mbox{\tiny{CGLMP}}} \rangle, \langle \mathcal{B}^{yz}_{\mbox{\tiny{CGLMP}}} \rangle, \langle \mathcal{B}^{zx}_{\mbox{\tiny{CGLMP}}} \rangle) \mbox{,}
\end{equation}
and this inequality is violated when:
\begin{equation}
    \mathcal{I}^{xyz}_{3} > 2\mbox{.}
\end{equation}
The operators are themselves constructed from the products of cosines of spin analysers of the parent $W$ boson (charged leptons and strange-type jets in this topology). For example, $\mathcal{B}^{xy}_{\mbox{\tiny{CGLMP}}}$, following the formalism in \cite{Barr:2021zcp} is formed from the summation of the averages of three observables:
\begin{align*}
    &\mathcal{O'}_{xy}^{1} = \tfrac{8}{\sqrt{3}} \langle {O}_{xy}^{1} \rangle \\
    &{O}_{xy}^{1} =  \xi_{x}^{+}\xi_{x}^{-} + \xi_{y}^{+}\xi_{y}^{-} \\[6pt]
    &\mathcal{O'}_{xy}^{2} = 25\langle {O}_{xy}^{2} \rangle\\
    &{O}_{xy}^{2} = ((\xi_{x}^{+})^2 - (\xi_{y}^{+})^2)((\xi_{x}^{-})^2 - (\xi_{y}^{-})^2)\\[6pt]
    &\mathcal{O'}_{xy}^{3} = 100\langle {O}_{xy}^{3} \rangle \mbox{,}\\
    &{O}_{xy}^{3} = \xi_{x}^{+}\xi_{y}^{+}\xi_{x}^{-}\xi_{y}^{-}\mbox{,}
\end{align*}
where $\xi_{xyz}^{\pm}$ is the cosine of the angle between the spin analyser from the parent $W^{\pm}$ boson and the spin analysing axis $\hat{x}\hat{y}\hat{z}$. Similar sets of three observables exist for the $\mathcal{B}^{yz}_{\mbox{\tiny{CGLMP}}}$ and $\mathcal{B}^{xz}_{\mbox{\tiny{CGLMP}}}$ operators. In each event the sign of the two $W$ is determined by the sign of the charged lepton. Each of the observables is constructed event-by-event and used to fill histograms from which the means are later determined. The means are then used to construct the relevant operator and, hence, determine the value for the inequality. 

\begin{figure}
     \centering
     \includegraphics[width=0.4\textwidth]{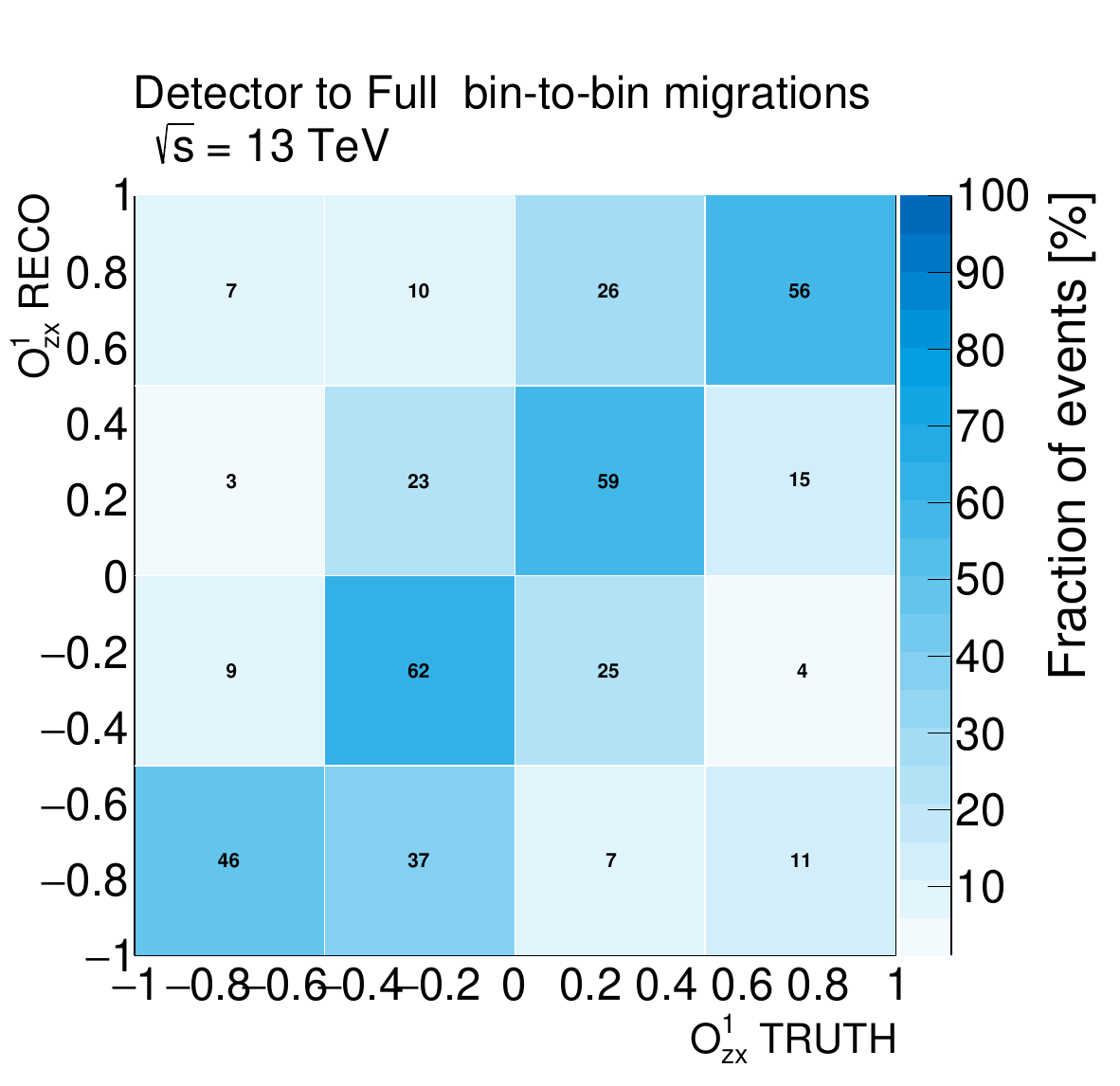}
     \caption{Migration matrices for one of the three observables composing the expectation value of the $zx$ Bell Operator.}
     \label{fig:migration}
\end{figure}

\subsection{Unfolding}

In order to correct for detector effects, many collider measurements use a technique called unfolding. We replicate this technique to remove the smearing effects imposed by our identification, efficiency, and reconstruction process using a type of unfolding called Iterative Bayesian Unfolding~\cite{DAGOSTINI1995487}. In particular, we use the procedure described in several ATLAS top quark measurements~\cite{ATLAS_top_dilep,ATLAS_top_ljets} and implemented in the \textsc{RooUnfold} package~\cite{roounfold}. The nine observables of interest are unfolded using two iterations and extrapolated to the full semi-leptonic phase space, at which point the mean of each observable is extracted to determine $\langle \mathcal{B}^{xy}_{\mbox{\tiny{CGLMP}}} \rangle, \langle \mathcal{B}^{yz}_{\mbox{\tiny{CGLMP}}} \rangle$, and $\langle \mathcal{B}^{zx}_{\mbox{\tiny{CGLMP}}} \rangle$. With the choice of axes and selection described in Section~\ref{sec:event_selection}, $\langle \mathcal{B}^{zx}_{\mbox{\tiny{CGLMP}}} \rangle$ is always the largest and thus is the value taken to determine $\mathcal{I}^{xyz}_{3}$. The normalised migration matrices for one of the observables used to build this operator is shown in Fig.~\ref{fig:migration}. The chosen binning results in diagonal matrices and consequently a stable unfolding. The limitations of the reconstruction are also apparent and manifest as a smearing of the reconstructed quantity towards zero. 

\begin{figure}[h!]
     \centering
     \hfill
     \includegraphics[width=0.48\textwidth]{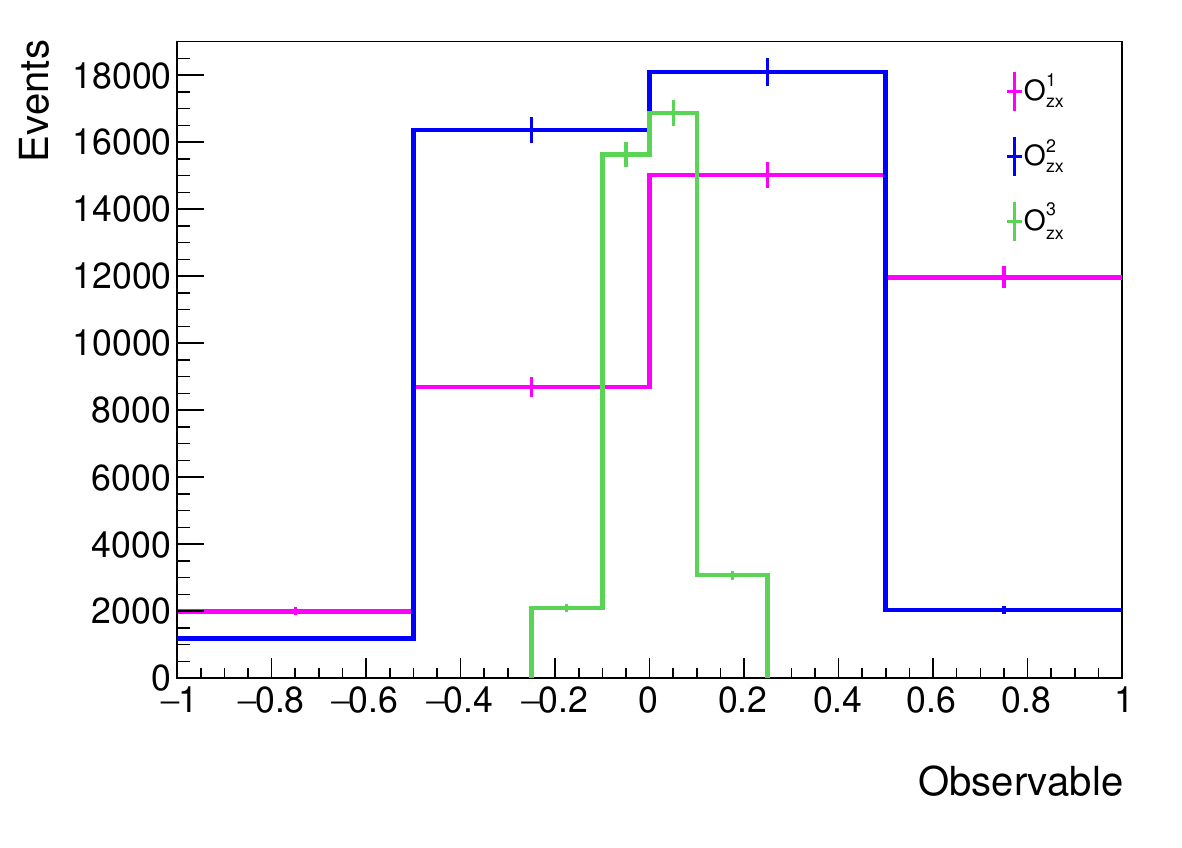}
     \caption{Unfolded histograms for the three observables composing the expectation value of the $zx$ Bell Operator. The uncertainties on each bin are statistical.}
     \label{fig:unfolded}
\end{figure}

The unfolded observables employed to determine $\langle \mathcal{B}^{zx}_{\mbox{\tiny{CGLMP}}} \rangle$ are shown in Fig.~\ref{fig:unfolded}.

\FloatBarrier
\subsection{Expected precision}

\begin{table*}
\caption{The expected sensitivity for different values of integrated luminosity for a realistic and idealised case. The uncertainties are statistical only.}
\label{tab:sensitivity}
\centering
\begin{tabular}{c r c l c}
\toprule
Luminosity [fb$^{-1}$]  &\multicolumn{3}{c}{$\langle \mathcal{B}^{zx}_{\mbox{\tiny{CGLMP}}} \rangle$} (idealised) & Significance (idealised) \\ 
\midrule
 139 & 2.45 & $\pm$ & 0.25 (0.18) & 1.8 (2.5) \\
 300 & 2.45 & $\pm$ & 0.17 (0.12) & 2.65 (3.75) \\
3000 & 2.45 & $\pm$ & 0.05 (0.04) & 9.0 (11.25) \\
\bottomrule
\end{tabular}
\end{table*}

Using the event selection, reconstruction, and unfolding described in the previous sections, the expected measurement and sensitivity for $\mathcal{I}_{3}$ is determined for three different values of integrated luminosity at a centre of mass energy of 13 TeV: 139 fb$^{-1}$ (corresponding to the amount of data collected by both ATLAS and CMS during LHC Run2), 300 fb$^{-1}$ (corresponding to the expected data that will be collected by the combination of Run2 and Run3), and 3000 fb$^{-1}$ (corresponding to the expected total integrated luminosity for HL-LHC). In all cases, the centre of mass energy is 13 TeV though, in practice, the LHC Run3 centre-of-mass energy is slightly higher and the energy for HL-LHC may reach 14 TeV. This small changes in the center of mass energy does not largely effect the numbers projected here, as the signal and background cross-sections are only mildly affected. Moreover it is possible to combine the results on the significance across the different center of mass energies. 
The uncertainties considered are only statistical, these are evaluated taking into account the unfolding procedure. The statistical uncertainty is estimated using pseudo-experiments, derived smearing every bin of the input distributions with a Poisson distribution. The standard deviation of the unfolded pseudo-experiments in each bin is taken as statistical uncertainty and then propagated to the histograms mean and to $\mathcal{I}^{xyz}_{3}$.

The expected measured values and significances for each of these luminosity scenarios are
presented in Table~\ref{tab:sensitivity}. The central value is 2.45, indicating violation of the $\mathcal{I}_{3}$ inequality. The expected uncertainties and significances of these values are presented for the charm tagging efficiency described in Section~\ref{sec:charm_tagging} and for an idealised case. Though such a case is not realistic for the existing LHC experiments, it provides a useful upper-bound on the results. For 300 $\mbox{fb}^{-1}$ the expected significance is slightly below 3 sigmas in the realistic case, and above it in the idealised case, suggesting that the evidence could be reached by the LHC experiments already at the end of Run3. In the HL-LHC scenario the expected significance rises to 9 sigmas in the realistic scenario and to 11 in the idealised one. The impact of the assumption of perfect jet reconstruction was tested here using performances derived from ATLAS public jet performance results~\cite{ATLAS_jet_smearing} for the case with non-idealised charm tagging, where the significances are slightly diluted to 1.6, 2.4, and 7.5 sigma for 130 $\mbox{fb}^{-1}$, 300 $\mbox{fb}^{-1}$,and 3000 $\mbox{fb}^{-1}$, respectively.

\section{Conclusions}

In this paper, we have demonstrated a method for isolating a previously inaccessible Higgs decay mode. The adaptation of Neutrino Weighting to semi-leptonic \hww final state provides a new tool to separate the signal from the overwhelming background allowing to reach a signal purity of 18\%, in combination with $c$-tagging and $b$-tagging. We have implemented a simple selection on the score of the NW, but the separation between signal and background could be further enhanced by introducing a multivariate analysis technique that combines the NW with other variables sensitive to the differences between \hww and $W$+jets.
Using the NW and the $c$-tagging combined with the $W$ and Higgs mass constraints, we have fully reconstructed the final state and prototyped a measurement of the CGLMP Bell-type inequality in this topology. The results are promising and a significance of almost 3 sigma is expected with the soon-to-be collected luminosity of 300$fb^{-1}$ at the LHC, whilst observation could be reached using the data sample expected for HL-LHC. We found that the reconstruction of the off-shell leptonic $W$ boson and the identification of the $c$-quark (and hence, $s$-quark) type jets are limiting factors for the measurement. Innovative charm tagging, with increased efficiency and an improved mis-tag rate for $b$-jets would represent a large improvement on the result both in terms of reduction of the statistical uncertainty and in the increase of the signal purity. This kind of improvement seems feasible in the near future, given recent advancements in flavour tagging algorithms~\cite{CMS:2021scf,ATL-PHYS-PUB-2022-047}. We focused only on the CGLMP inequalities as a probe to study the quantum nature of the Higgs boson but the same approach could be applied to measure other entanglement witnesses valid for qutrit systems introduced in recent phenomenological studies~\cite{Aguilar-Saavedra:2022wam}. This could facilitate observation of entanglement in \hww events before reaching the statistic required to observe the violation of Bell inequalities.

\clearpage

\printbibliography
\end{document}